\documentclass[journal,12pt,draftclsnofoot,onecolumn]{IEEEtran}
\IEEEoverridecommandlockouts
\ifCLASSINFOpdf
\else
\fi
\usepackage[cmex10]{amsmath}
\usepackage{amsthm,cite,graphicx,,booktabs,lipsum,color,bm,algorithm,caption,subcaption}
\usepackage{algpseudocode,pifont,tikz,paralist,multirow,amssymb}
\usepackage[labelformat=simple]{subcaption}

\theoremstyle{definition}

\newtheorem{lem}{Lemma}
\begin{document}

\begin{titlepage}
\begin{center}
\vspace*{-2\baselineskip}
\begin{minipage}[l]{7cm}
\flushleft
\includegraphics[width=2 in]{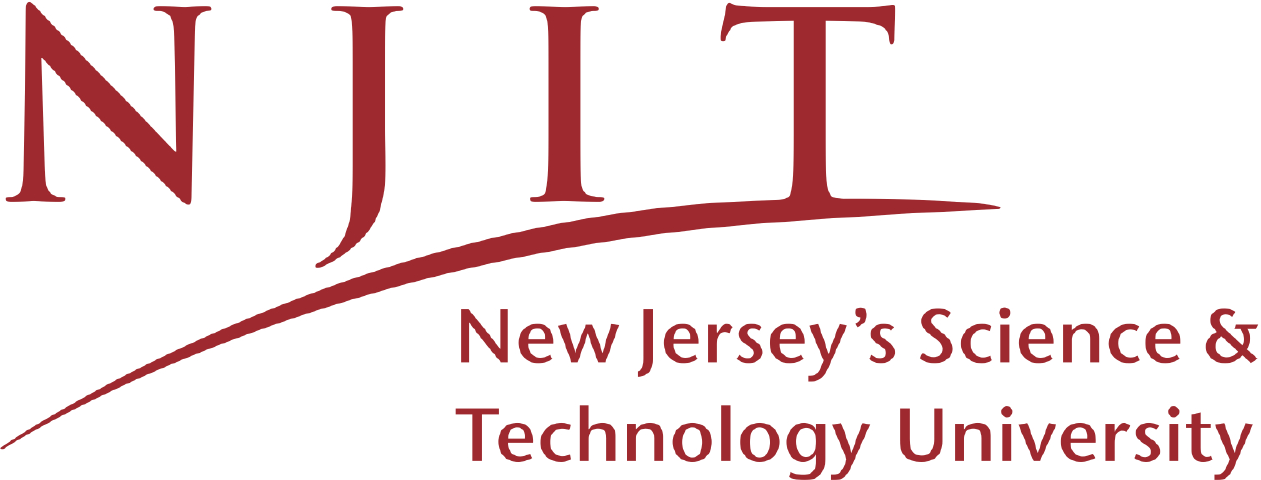}
\end{minipage}
\hfill
\begin{minipage}[r]{7cm}
\flushright
\includegraphics[width=1 in]{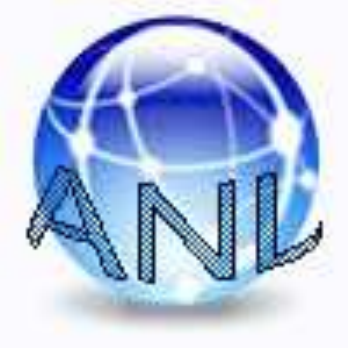}
\end{minipage}

\vfill

\textsc{\LARGE Latency Aware Drone Base Station Placement in Heterogeneous Networks\\[12pt]}
\vfill
\textsc{
\LARGE  XIANG SUN \\ NIRWAN ANSARI}\\
\vfill
\textsc{\LARGE TR-ANL-2017-004\\[12pt]
\LARGE June 1, 2017}\\[1.5cm]

\vfill
{ADVANCED NETWORKING LABORATORY\\
 DEPARTMENT OF ELECTRICAL AND COMPUTER ENGINEERING\\
 NEW JERSY INSTITUTE OF TECHNOLOGY}
\end{center}
\end{titlepage}
\title{Latency Aware Drone Base Station Placement in Heterogeneous Networks}

\author{Xiang~Sun,~\IEEEmembership{Student~Member,~IEEE,}
        Nirwan~Ansari,~\IEEEmembership{Fellow,~IEEE} 
\thanks{X. Sun and N. Ansari are with Advanced Networking Lab., Department of Electrical $\&$ Computer Engineering, New Jersey Institute of Technology, Newark, NJ 07102, USA. E-mail:$\{$xs47, nirwan.ansari$\}$@njit.edu.
\textcolor{blue}{\bf{The paper is going to be presented at IEEE GLOBCOM 2017}}.}
}

\maketitle

\begin{abstract}
Different from traditional static small cells, Drone Base Stations (DBSs) exhibit their own advantages, i.e., faster and cheaper to deploy, more flexibly reconfigured, and likely to have better communications channels owing to the presence of short-range line-of-sight links. Thus, applying DBSs into the cellular network has great potential to increase the throughput of the network and improve Quality of Service (QoS) of Mobile Users (MUs). In this paper, we focus on how to place the DBS (i.e., jointly determining the location and the association coverage of a DBS) in order to improve the QoS in terms of minimizing the total average latency ratio of MUs by considering the energy capacity limitation of the DBS. We formulate the DBS placement problem as an optimization problem and design a Latency aware dronE bAse station Placement (LEAP) algorithm to solve it efficiently. The performance of LEAP is demonstrated via simulations as compared to other two baseline methods.
\end{abstract}

\IEEEpeerreviewmaketitle

\section{Introduction}

A drone is considered as an Unmanned Aerial Vehicle (UAV) that is designed to be flown under remote control or autonomously using embedded software and sensors (e.g., GPS) \cite{Sedjelmaci:IDE:2017}. Recently, drones have been incorporated into the cellular system to facilitate communications between Mobile Users (MUs) and the cellular network \cite{Sun:MEC:2017}. For instance, Nokia has deployed consumer quadcopter drones armed with pico cells in order to expand the mobile network coverage in rural areas of the UK. The designed Drone Base Station (DBS) can provide the coverage over a radius of 5 $km$, enabling high quality voice calls between the MUs, real-time video streaming and up to 150 $Mbps$ data throughput \cite{Nokia}. DBS can provide nearly the same performance as compared to traditional small base stations, which are deployed in urban areas to boost network throughput, improve Quality of Service (QoS) of MUs, and increase the energy efficiency of the network. However, different from traditional small base stations, whose locations are normally fixed, DBSs provision flexible small cell deployment, which has the potential to improve the network performance. Specifically, the traffic demands from MUs exhibits spatial and temporal dynamics, and thus the static small base station deployment may not always be the optimal solution to meet the traffic demands from MUs. The spatial and temporal dynamic features of the traffic loads require the flexible small cell deployment, which enables DBSs to be a promising and economic solution to improve QoS of MUs. 

As compared to static small cells, DBSs have their own drawbacks, i.e., DBSs are powered by batteries, and thus cannot serve the corresponding MUs continuously. DBSs should return back to the charging station before their batteries are exhausted. This requires the system to carefully schedule the energy consumption of the DBSs to avoid their crashing off. 

\begin{figure}[!htb]
	\centering	
	\includegraphics[width=0.8\columnwidth]{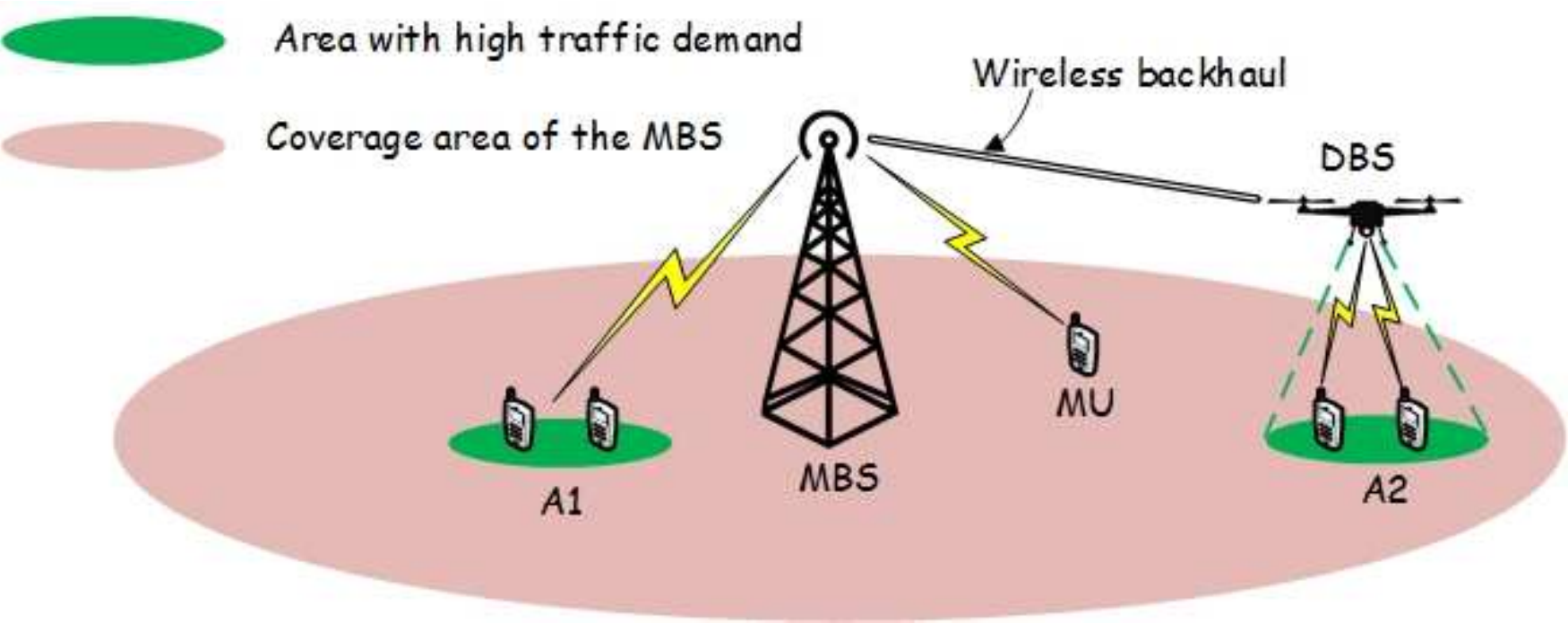}
	\caption{The DBS enabled heterogeneous network.}	
	\label{fig:scenario}
\end{figure}

In this paper, we consider one Macro Base Station (MBS) and one DBS that coexist in the network. As shown in Fig. \ref{fig:scenario}, the DBS can offload some traffic loads from the MBS via the wireless backhaul and deliver them to the MUs. However, it is still unclear on how to efficiently place the DBS, which can be further divided into the following two problems:
\begin{enumerate}
\item{Where should the DBS be deployed?\\}
{Normally, the DBS can be deployed in an area with a high traffic demand to provide a high data rate \cite{Sharma:2016:UAH}. However, the traffic demand is not the only factor to determine the DBS deployment. The channel condition between the area and the MBS can also affect the DBS deployment. For example, consider two areas with high traffic demands within the MBS's coverage, i.e., A1 and A2. A1 is close to the MBS and achieves the better channel condition, while A2 is at the edge of the MBS's coverage and suffers from the worse channel condition with respect to the MBS. Obviously, deploying the DBS at A2 provides better QoS to MUs. Thus, the we should jointly consider the traffic demands and the channel conditions among the areas when placing the DBS.}
\item{What is the association coverage of the DBS?\\}
{The association coverage of the DBS refers to the set of MUs which can receive (download) data from the DBS. The association coverage of the DBS determines the traffic loads of the DBS, i.e., if the DBS has a larger association coverage, the DBS would deliver more data to the corresponding MUs that may significantly increase the delay as well as the energy consumption of the DBS. Thus, it is necessary to balance the traffic loads between the DBS and the MBS to achieve the minimum delay in delivering data to the MUs while meeting the energy capacity limitations of the DBS.}
\end{enumerate}
Note that the two problems are coupled together, i.e., deploying the DBS in a different location may result in a different association coverage of the DBS.

\section{Related works}
As compared to the traditional terrestrial wireless communications \cite{Sun:2015:EBA}, drone-aided wireless systems have been identified with their unique advantages, i.e., faster and cheaper to deploy, more flexibly reconfigured, and likely to have better communications channels owing to the presence of short-range line-of-sight links \cite{Zeng:2016:WCU}. Some works have been done on how to place/deploy DBSs in the network. Al-Hourani \emph{et al.} \cite{AlHourani:2014:OLA} derived the optimal altitude of a DBS, which is a function of the maximum allowed pathloss and the statistical parameters of the urban environment, to maximize the radio coverage on the ground. Mozaffari \emph{et al.} \cite{Mozaffari:2015:DSC} investigated the problem of placing two DBSs. They derived the optimal altitude of the two DBSs as well as the optimal distance between the two DBSs to maximize the total coverage area. Yaliniz \emph{et al.} \cite{Yaliniz:2016:EPA} proposed a DBS placement problem to determine the location, the altitude as well as the coverage area of a DBS such that the DBS can cover as many MUs as possible. Azade \emph{et al.} \cite{Fotouhi:2016:DBS} designed a DBS repositioning method to move the DBS to the direction, which achieves the largest spectral efficiency gain during a time slot. Here, the spectral efficiency gain indicates the average spectral efficiency difference between the DBS and the MBS for serving the MUs.

Different from the above works, we try to design an optimal DBS placement (i.e., jointly optimizing the location and the association coverage of the DBS) to improve the QoS of MUs by considering the energy constraint of the DBS. 

\section {System model and problem formulation}
The application scenario is depicted in Fig. \ref{fig:scenario}, where one DBS assists the MBS to serve the corresponding MUs.

\subsection{Traffic model of the MBS}
The whole coverage area is divided into a number of locations with the same size. Denote $\bm{\mathcal{I}}$ as the set of these locations and $i$ is used to index these locations. We assume that the traffic arrives according to a Poisson point process with the average arrival rate per unit area at location $i$ equaling to $\lambda_i$, and the traffic size (packet size) per arrival has a general distribution with the average traffic size of $\nu_i$. Thus, the average traffic load at location $i$ is $\lambda_i\nu_i$.

If an MU at location $i$ is associated with the MBS, then the MU's data rate, denoted as $r^m_i$, can be expressed as
\begin{equation}
r_i^m = {w^m}{\log _2}\left( {1 + \frac{{{P^m}g_i^m}}{{{\sigma ^2+\iota_i^m}}}} \right),
\label{eq:MBS_data_rate}
\end{equation}
where $w^m$ is the total amount of bandwidth available for the MBS, $P^m$ is the transmission power of the MBS, $\sigma ^2$ denotes the noise power level, $\iota^m_i$ is the average interference power seen by an MU at location $i$ from other MBSs, and $g_i^m$ is the channel gain between the MBS and an MU at location $i$. Thus, $\frac{{{P^m}g_i^m}}{{{\sigma ^2+\iota_i^m}}}$ implies the SINR at location $i$. Note that $g_i^m$ can be measured by the MBS at large time scale and the value of $w^m$ is given based on the network's frequency allocation strategy. Consequently, the average utilization of the MBS in delivering the traffic loads to the MUs in location $i$ is
\begin{equation}
\rho _i^m = \frac{{{\lambda _i}{\nu _i}}}{{{r^m_i}}}.
\end{equation}
The value of $\rho _i^m$ indicates the fraction of time during which the MBS is busy serving the MUs in location $i$. Thus, we can derive the average utilization of the MBS in delivering the traffic loads to the MUs in the MBS's association coverage as
\begin{equation}
{\rho ^m} = \sum\limits_{i \in \bm{\mathcal{I}}} {\rho _i^m\left({1-\theta_i}\right)}  = \sum\limits_{i \in \bm{\mathcal{I}}} {\frac{{{\lambda _i}{\nu _i}\left({1-\theta_i}\right)}}{{{r^m_i}}}},
\label{eq:mbs_utilization}
\end{equation}
where $\theta_i$ is a binary variable indicating the location association strategy, i.e., if all the MUs in location $i$ are associated with MBS, then $\theta_i=0$; otherwise (i.e., the MUs in location $i$ are associated with DBS), $\theta_i=1$. Note that the value of $\rho^m$ indicates the fraction of time during which the MBS is busy serving its associated MUs. 

We assume that traffic arrivals at different locations are independent. Since the traffic arrival per location is a Poisson point process, the traffic arrival in the MBS, which is the sum of the traffic arrivals in its associated locations, is also a Poisson point process. The required service time per traffic arrival at the MBS's associated location $i$ is $s^m_i=\frac{{{\nu _i}}}{{{r^m_i}}}$, where ${\nu _i}$ is the average traffic size per arrival which follows a general distribution, and thus the required service time is also a general distribution. Thus, the MBS's downlink transmission process follows an M/G/1 processor sharing queue, in which multiple MUs within the MBS's association coverage share the MBS's downlink radio resource. According to \cite{Kleinrock:1976:QSC}, the average traffic delivery time, including waiting time and service time, for the MU at the MBS's association coverage $i$ is 
\begin{equation}
T_i^m = \frac{{s_i^m}}{{1 - {\rho ^m}}}.
\end{equation}

Denote $\tau^m_i$ as the average latency ratio that measures how much time an MU at the MBS's associated location $i$ must be sacrificed in waiting for a unit service time, i.e.,
\begin{equation}
\tau _i^m = \frac{{T_i^m - s_i^m}}{{s_i^m}} = \frac{{{\rho ^m}}}{{1 - {\rho ^m}}}.
\end{equation}
It is worth noting that the value of $\tau _i^m$ only depends on the the utilization of the MBS (i.e., $\rho ^m$). Therefore, the MUs in different locations (which are associated with the MBS) would have the same average latency ratio, i.e.,
\begin{equation}
\tau^m = \frac{{{\rho ^m}}}{{1 - {\rho ^m}}}.
\end{equation}
Note that we consider the average latency ratio of an MU as a metric to measure the QoS of the MU \cite{Han:2016:TLB}. A smaller value of the average latency ratio implies that the MU suffers from less average waiting time before they are served.

\subsection{Traffic model of the DBS}
By applying the similar derivation, we can obtain the average latency ratio of the MUs (denoted as $\tau ^d$), which are associated with the DBS, as
\begin{equation}
{\tau ^d} = \frac{{{\rho ^d}}}{{1 - {\rho ^d}}},
\end{equation}
where $\rho ^d$ is the average utilization of the DBS in delivering the traffic loads to MUs in the DBS's associated locations, i.e., 
\begin{equation}
{\rho ^d} = \sum\limits_{i \in \bm{\mathcal{I}}} {\rho _i^d\theta_i}  = \sum\limits_{i \in \bm{\mathcal{I}}} {\frac{{{\lambda _i}{\nu _i}\theta_i}}{{r_{ij}^d}}}.
\label{eq:dbs_utilization}
\end{equation}
Note that $\rho _i^d$ indicates the fraction of time during which the DBS is busy delivering the traffic to the MUs in location $i$ and ${r_{ij}^d}$ is the data rate of the MU at location $i$ in downloading the traffic from the DBS at location $j$. Thus, we have
\begin{equation}
r_{ij}^d = {w^d}{\log _2}\left( {1 + \frac{{{P^d}g_{ij}^d}}{{{\sigma ^2+\iota_i^d}}}} \right),
\label{eq:dbs_datarate}
\end{equation}
where $w^d$ is the total amount of bandwidth available for the DBS, $P^d$ is the transmission power of the DBS and $g_{ij}^d$ is the channel gain between the DBS at location $j$ ($j \in \bm{\mathcal{I}}$) and an MU at location $i$. Assume that $g_{ij}^d$ is mainly determined by the path loss (in dB) between the DBS at location $j$ and the MU at location $i$ (i.e., $g_{ij}^d={10^{\frac{{-\eta _{ij}^d}}{{10}}}}$, where $\eta_{ij}^d$ is the path loss), which can be modeled as
\begin{equation}
\eta _{ij}^d = \alpha  + \gamma {\log _{10}}\left( {{d_{ij}}} \right),
\end{equation}
where $\alpha$ is the path loss at the reference distance and $\gamma$ is the path loss exponent, both of which can be obtained from field tests \cite{3GPP:2012:release_11}. and $d_{ij}$ is the distance between the DBS at location $j$ and the MU at location $i$, i.e.,
\begin{equation}
{d_{ij}} = \sqrt {{{\left( {{x_i} - {x_j}} \right)}^2} + {{\left( {{y_i} - {y_j}} \right)}^2} + h^2},
\label{eq:distance}
\end{equation}
where $\left\langle {{x_i},{y_i}} \right\rangle$ and $\left\langle {{x_j},{y_j}} \right\rangle$ are the coordinations of location $i$ and $j$, respectively; $h$ is the DBS's height\footnote{Note that we assume the DBS always keeps the same height.}.

Consequently, we can obtain the data rate of an MU (when the MU is at location $i$ and the DBS at location $j$) as
\begin{equation}
r_{ij}^d\!=\!{w^d}{\log _2}\left( {1\!+\!\frac{{{P^d}{{10}^{\frac{{ -\alpha-10\gamma {\log _{10}}d_{ij} }}{{10}}}}}}{{{\sigma ^2}}}} \right).
\end{equation}


\subsection{Energy consumption of the DBS}

\begin{figure}[!htb]
	\centering	
	\includegraphics[width=1.0\columnwidth]{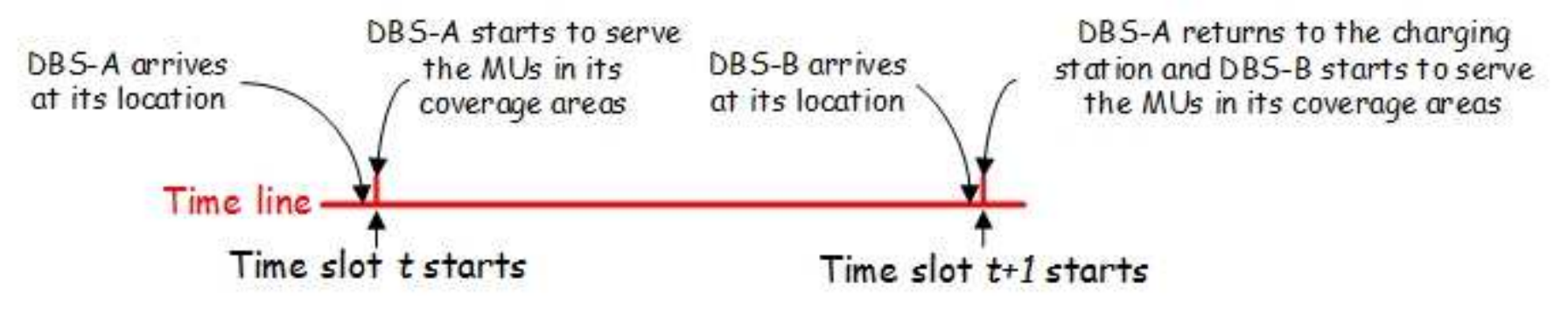}
	\caption{The DBS scheduling over time.}	
	\label{fig:DBS_scheduling}
\end{figure}

In each time slot, there is only one DBS that is running in its working state to serve the MUs within the DBS's coverage area. For example, as shown in Fig. \ref{fig:DBS_scheduling}, if DBS-A is determined to help the MBS to serve some of the MUs in time slot $t$, then DBS-A should arrive at the corresponding location before time slot $t$, starts to serve the MUs within its coverage areas as time slot $t$ begins, and returns back to the MBS for charging its battery as time slot $t+1$ begins. Note that once time slot $t+1$ starts, another DBS (e.g., DBS-B) would start to serve the MUs within its coverage areas.

DBS is considered as a type of small cell, whose power consumption is proportional to its traffic loads in terms of the utilization of the small cell. Thus, we model the power consumption of the DBS as \cite{Han:2016:TLB}
\begin{equation}
p = \beta {\rho ^d} + {p^s},
\end{equation}
where $\beta$ is the load-power coefficient that maps the utilization of the DBS into the power consumption and $p^s$ is the static power consumption of the DBS\footnote{The static power consumption of a DBS comprises two parts: the static power consumption from the communications module of the DBS (i.e., the power consumption of the communications module when $\rho^d=0$) and the power consumption of the flight module of the DBS (i.e., the power consumption of the DBS hovering at height $h$).}. Thus, in order to guarantee the DBS can fly to the location, serve the corresponding MUs in its association coverage for a time slot, and return back to the MBS for charing the battery, we have:
\begin{equation}
\Delta T\left( {\beta {\rho ^d} + {p^s}} \right) \le \varepsilon,
\end{equation}
where $\Delta T$ is the length of one time slot and $\varepsilon$ is a predefined threshold. Note that the value of $\varepsilon$ should be larger than the total battery capacity of the DBS minus the energy consumption of the DBS flying from the MBS to the edge of the MBS and back to the MBS (for charging the battery). 

\subsection{Problem formulation}
We formulate the problem as follows:
\begin{align}
	\bm{P0}:\ 
	&\mathop {argmin }\limits_{j,\bm{\theta} } {\frac{{{\rho ^m}}}{{1 - {\rho ^m}}} +\frac{{{\rho ^d}}}{{1 - {\rho ^d}}}}\\ 
	s.t.\ \ \ \  \ &{\rho ^m} = \sum\limits_{i \in \bm{\mathcal{I}}} {\frac{{{\lambda _i}{\nu _i}{\left(1-\theta _i\right)}}}{r^m_i}}, \label{ct1}\\
	&{\rho ^d}\!=\!\sum\limits_{i \in \bm{\mathcal{I}}} {\frac{{{\lambda _i}{\nu _i}{\theta _i}}}{r^{d}_{ij}}}, \label{ct2}\\
	&{\rho ^d} \le \frac{1}{\beta }\left( {\frac{\varepsilon }{{\Delta T}} - {p^s}} \right), \label{ct3}\\
	&\forall i \in {\bm{\mathcal{I}}}, \theta_i \in \left\{ {0,1} \right\} \label{ct4},\\
	& 0 \le \rho ^m, \rho ^d < 1, \label{ct5} 
\end{align}	 
where $j$ is the location of the DBS and $\bm{\theta}=\left\{ {{\theta _i}\left| {i \in \bm{\mathcal{I}}} \right.} \right\}$ is the location association vector. The objective is to minimize the total average latency ratio incurred by the DBS and the MBS. Constraints (\ref{ct1}) and (\ref{ct2}) are the utilization of the MBS and the DBS, respectively. Constraint (\ref{ct3}) indicates that the energy consumption of the DBS during the time slot should be less than the predefined threshold. Constraint (\ref{ct4}) implies that $\theta _i$ is a binary variable. Constraint (\ref{ct5}) implies that the utilization of the MBS and DBS should be between 0 and 1.

\section{Latency aware dronE bAse station Placement}

We will introduce the LEAP algorithm to solve $\bm{P0}$. Basically, LEAP is to first determine the location of the DBS and then optimize the association coverage of the DBS.

Assume $\rho^m+\rho^d=\rho$; then $\bm{P0}$ can be transformed into
\begin{align}
	\bm{P1}:\ 
	& \mathop {argmin }\limits_{\rho ^m, \rho ^d} {\frac{{{\rho ^m}}}{{1 - {\rho ^m}}} +\frac{{{\rho ^d}}}{{1 - {\rho ^d}}}} \nonumber\\
	s.t.\ \ \ \  \ &\rho^m+\rho^d=\rho, \nonumber\\
	&{\rho ^d} \le \frac{1}{\beta }\left( {\frac{\varepsilon }{{\Delta T}} - {p ^s}} \right), \nonumber\\
	&{\rho ^m},{\rho ^d}\ge 0. \nonumber
\end{align}
	 
It is easy to derive $\bm{P1}$ to be a convex problem. By applying the Karush Kuhn Tucker (KKT) conditions, we can obtain the close form optimal solutions of $\bm{P1}$, i.e., ${\rho ^d}=\min \left\{ {\frac{\rho }{2},\frac{1}{\beta }\left( {\frac{\varepsilon }{{\Delta T}}\!-\!{p^s}} \right)} \right\}$ and ${\rho ^m}=\max \left\{ {\frac{\rho }{2},\rho \!-\!\frac{1}{\beta }\left( {\frac{\varepsilon }{{\Delta T}}\!-\!{p ^s}} \right)}\right\}$.

\subsection{Optimal location of the DBS}
In this section, we will design a method to determine the DBS's location to minimizing the total latency ratio. 

Note that minimizing the value of $\rho$ (where $\rho=\rho^m+\rho^d$) is equivalent to minimizing the total latency ratio, and thus we will try to find the optimal location $j^*$ ($j^* \in \bm{\mathcal{I}}$) such that the value of $\rho$ is the minimum.

\begin{lem}
The optimal location $j^*$ can be derived from
\begin{equation}
{j^*} = \mathop {\arg \min }\limits_j \left\{ {\sum\limits_{i \in {\bm{\mathcal{I}}_j}} {{\lambda _i}{\nu _i}\left( {\frac{1}{{r_{ij}^d}} - \frac{1}{{r_i^m}}} \right)}} \right\},
\label{eq:opt_location}
\end{equation}
where ${\bm{\mathcal{I}}_j} = \left\{ {i \in {\bm{\mathcal{I}}}\left| {r_{ij}^d \ge r_i^m} \right.} \right\}$.
\label{lem:obj_location}
\end{lem} 

\begin{IEEEproof}
\begin{align}
\rho &=\sum\limits_{i \in \bm{\mathcal{I}}} {{\lambda _i}{\nu _i}\left( {\frac{1}{{r_{ij}^d}} - \frac{1}{{r_i^m}}} \right){\theta _i}}  + \sum\limits_{i \in \bm{\mathcal{I}}} {\frac{{{\lambda _i}{\nu _i}}}{{r_i^m}}}.
\label{eq:value_of_rho}
\end{align}

Thus, if the DBS is placed at location $j$, in order to minimize $\rho$, $\theta _i$ should equal to 1 (i.e., location $i$ is associated with the DBS) iff ${\frac{1}{{r_{ij}^d}} - \frac{1}{{r_i^m}}}<0$, i.e., $\forall i \in {\bm{\mathcal{I}}_j},{\theta _i} = 1$, where ${\bm{\mathcal{I}}_j} = \left\{ {i \in {\bm{\mathcal{I}}}\left| {r_{ij}^d > r_i^m} \right.} \right\}$. The physical meaning of ${\bm{\mathcal{I}}_j}$ is the set of locations, where the DBS (that is currently at location $j$) can provides higher data rate than the MBS. Consequently, the corresponding value of $\rho$ is
\begin{equation}
\rho=\sum\limits_{i \in {\bm{\mathcal{I}}_j}} {{\lambda _i}{\nu _i}\left( {\frac{1}{{r_{ij}^d}} - \frac{1}{{r_i^m}}} \right)}  + \sum\limits_{i \in {\bm{\mathcal{I}}}} {\frac{{{\lambda _i}{\nu _i}}}{{r_i^m}}}.
\label{eq:rho_per_location}
\end{equation}

Obviously, in order to find the optimal location to minimize the value of $\rho$, we should find the location, which incurs the minimum value of $\sum\limits_{i \in {\bm{\mathcal{I}}_j}} {{\lambda _i}{\nu _i}\left( {\frac{1}{{r_{ij}^d}} - \frac{1}{{r_i^m}}} \right)}$ among all the locations, i.e., ${j^*} = \mathop {\arg \min }\limits_j \left\{ {\sum\limits_{i \in {\bm{\mathcal{I}}_j}} {{\lambda _i}{\nu _i}\left( {\frac{1}{{r_{ij}^d}} - \frac{1}{{r_i^m}}} \right)}} \right\}$, where ${j^*}$ is the optimal location of the DBS.
\end{IEEEproof}

\subsection{Optimal coverage of the DBS}
In this section, we will design a method to find the optimal association coverage of the DBS (i.e., the value of $\bm{\theta}$) given the optimal location of the DBS, i.e., the value of $j^*$.

\begin{figure}[!htb]
	\centering	
	\includegraphics[width=0.6\columnwidth]{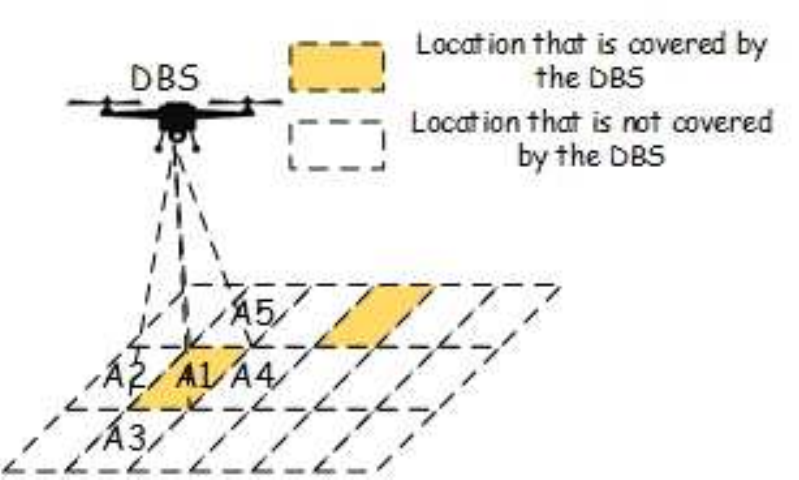}
	\caption{The illustration of neighbor location set and discontinuous coverage.}
	\label{fig:impossible_coverage}
\end{figure}

Assume all the locations are initially associated with the MBS. Since the optimal utilization of the DBS is $\rho^{d*}=\min \left\{ {\frac{\rho }{2},\frac{1}{\beta }\left( {\frac{\varepsilon }{{\Delta T}}\!-\!{p^s}} \right)} \right\}$, the basic idea of the method is to iteratively select the most suitable location from the neighbor location set and let it be associated with the DBS until the utilization of the DBS is larger than $\rho^{d*}$. Here, the neighbor location set refers to as all the locations (which are currently associated with the MBS) that are the neighbors of the DBS's associated locations (i.e., the locations that have already been associated with the DBS). For instance, as shown in Fig. \ref{fig:impossible_coverage}, if only location A1 is currently associated with the DBS, then A2, A3, A4, and A5 are the neighbor locations of A1. The reason for doing that is to avoid the discontinuous association coverage as illustrated in Fig. \ref{fig:impossible_coverage}. Note that the most suitable location is defined as the location, which is currently associated with the MBS, that incurs the minimum value of $\rho$ (if the location is associated with the DBS) among all the locations in the neighbor location set, i.e., 
\begin{equation}
{i^*} = \mathop {\arg \min }\limits_{i \in {{\widetilde {\bm{\mathcal{I}}}}_{{j^*}}}} \left\{  {{\lambda _i}{\nu _i}\left( {\frac{1}{{r_{ij^*}^d}} - \frac{1}{{r_i^m}}} \right)}  \right\},
\label{eq:suitable_location}
\end{equation}
where ${{\widetilde {\bm{\mathcal{I}}}}_{{j^*}}}$ denotes the set of the neighbor location set. Note that the neighbor location set and the value of $\rho$ (which is based on Eq. \eqref{eq:value_of_rho}) should be updated for each iteration.



\subsection{Summary of LEAP}
The basic idea of LEAP is to first determine the location of the DBS, and then optimize the coverage of the DBS. The LEAP algorithm is summarized in Algorithm \ref{alg:LEAP}.
	\begin{algorithm}
		\caption{The LEAP algorithm}
		\label{alg:LEAP}
		\begin{algorithmic}[1]
			\State Find the DBS's optimal location based on Lemma. \ref{lem:obj_location}.
			\State Initialize the location association vector $\bm{\theta}$, where ${\theta _{{j^*}}} = 1$ and ${\theta _{i \in \bm{\mathcal{I}}\backslash {j^*}}} = 0$.
			\State Obtain the neighbor location set ${{\widetilde {\bm{\mathcal{I}}}}_{{j^*}}}$.
			\State Calculate the current utilization of the DBS ${\rho ^d} = \frac{{{\lambda _{{j^*}}}{\nu _{{j^*}}}}}{{r_{{j^*}{j^*}}^d}}$.
			\State Calculate the value of $\rho$ based on Eq. \eqref{eq:value_of_rho}.
		    \State Find the current suitable location $i^*$ based on Eq. \eqref{eq:suitable_location}.
		    \State $\rho=\rho+{{\lambda _{i^*}}{\nu _{i^*}}\left( {\frac{1}{{r_{i^*j^*}^d}} - \frac{1}{{r_{i^*}^m}}} \right)}$.
			\While {${\rho ^d}+ \frac{{{\lambda _{{i^*}}}{\nu _{{i^*}}}}}{{r_{{i^*}{j^*}}^d}} < \min \left\{ {\frac{\rho }{2},\frac{1}{\beta }\left( {\frac{\varepsilon }{{\Delta T}}\!-\!{p^s}} \right)} \right\}$}
			\State Associate location $i^*$ with the DBS, i.e., ${\theta _{{i^*}}} = 1$;
			\State Update the utilization of the DBS ${\rho ^d}\!=\!{\rho ^d}\!+\!\frac{{{\lambda _{{i^*}}}{\nu _{{i^*}}}}}{{r_{{i^*}{j^*}}^d}}$.
			\State Update the neighbor location set ${{\widetilde {\bm{\mathcal{I}}}}_{{j^*}}}$;
			\State Find the current suitable location $i^*$ based on Eq. \eqref{eq:suitable_location};
			\State Update $\rho=\rho+{{\lambda _{i^*}}{\nu _{i^*}}\left( {\frac{1}{{r_{i^*j^*}^d}} - \frac{1}{{r_{i^*}^m}}} \right)}$.
			\EndWhile
		\end{algorithmic}
	\end{algorithm}

\section{Simulations}
We set up system level simulations to investigate the performance of LEAP. We apply the MU movement trace provided by the EveryWare Lab. The trace provides the MUs’ movement in the road network of Milan. The whole road network size is 17 $\times$ 28.64 $km$. There are a total of 100,000 MUs in the area. Parameters of the MU movement trace are detailed in \cite{user_trace}. In the simulation, we choose a 1 $\times$ 1 $km$ area of the whole road network and obtain the movements of MUs within this area from 7 pm to 1 am. The area is further divided into 10,000 locations with each location representing a 10 $\times$ 10 $m$ small area. An MBS is placed in the central of the area. The traffic arrivals for each MU follows a Poison distribution with the average traffic arrival rate equaling to 0.15 $request/s$ and the average traffic size per arrival is 100 $kb$. The height of the DBS is $h= 10\ m$ and the power consumption of a drone hovering in the air is 110 $watt$ \cite{Franco:2015:ECP}. The static power consumption of a small cell is 37 $watt$ \cite{Auer:2011:HME}. The load-power coefficient of the DBS is $\beta = 500$. The energy threshold $\epsilon = 0.2$ $kWh$. The total bandwidth is 20 $MHz$ in which 15 $MHz$ is exclusively used by MBSs and the other 5 $MHz$ is allocated to DBS. The other parameters are summarized in Table \ref{tab:para}.  

\begin{table}[htbp]
  \centering
  \caption{Values and Definitions of Parameters}
    \begin{tabular}{lll}
    \toprule
    \textbf{Parameters} &\textbf{Definition} & \textbf{Value} \\
    \midrule
    $P^m$  & Transmission power of the MBS & 46 $dBm$ \\
    $P^d$  & Transmission power of the DBS & 24 $dBm$ \\
    $PL_{MBS}$   & Path loss model of the MBS  & $103.4+2.42{\log _{10}}d$ \\
    $PL_{DBS}$ & Path loss model of the DBS    & $103.8+2.09{\log _{10}}d$ \\
    $\sigma ^2$& Noise power level 	       & 174 $dBm$ \\
    $\Delta T$ & Time slot duration        & 10 $min$ \\
    \bottomrule
    \end{tabular}%
  \label{tab:para}%
\end{table}%

\begin{figure*}[!htb]
\minipage{0.32\textwidth}
	\centering
	\begin{subfigure}{1.0\textwidth}
		\centering
		\includegraphics[width=1.0\linewidth]{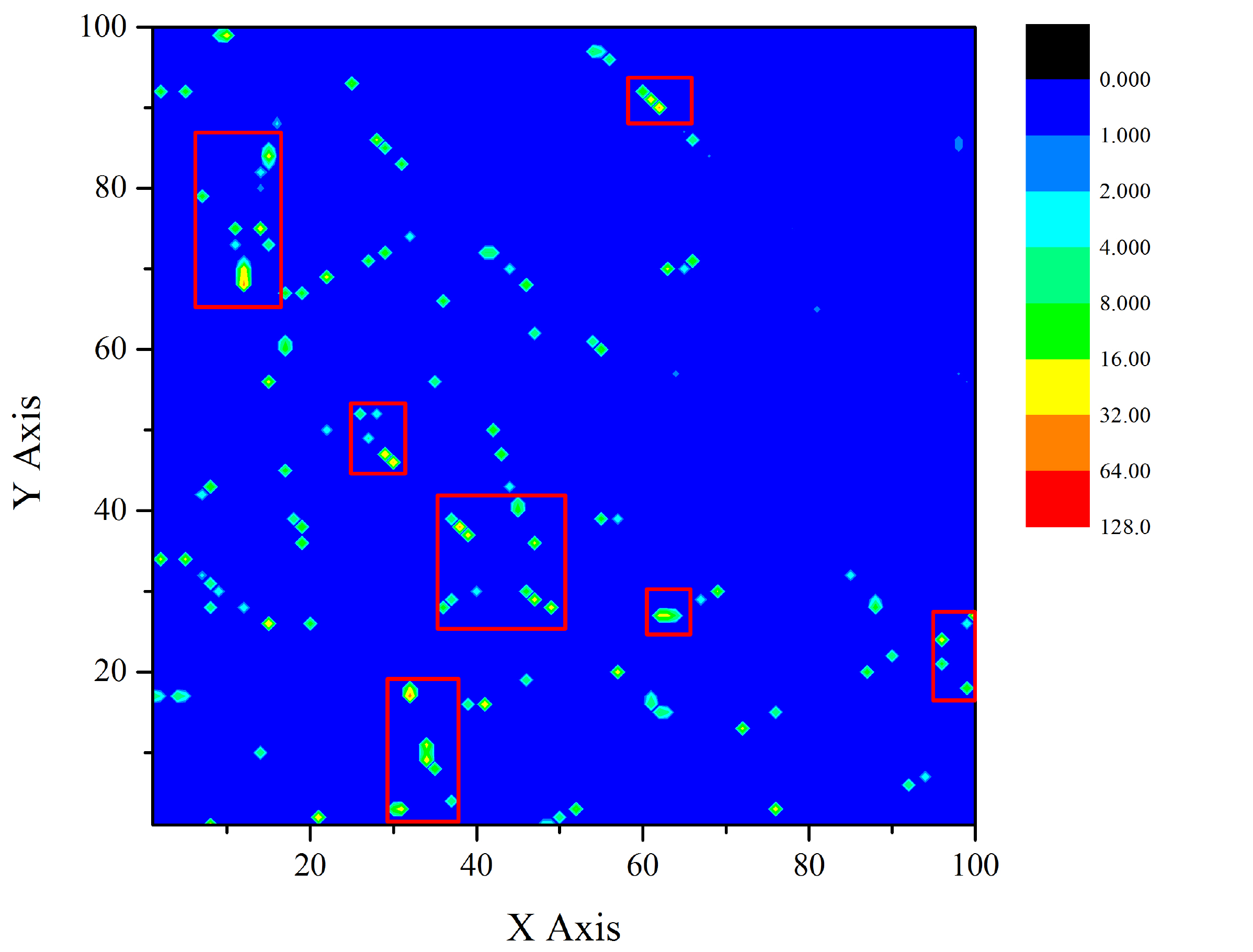}
		\caption{The MU density in the network when $t=1$.}
		\label{fig:MU_dens_1}
	\end{subfigure}%
	\\
	\begin{subfigure}{1.0\textwidth}
		\centering
		\includegraphics[width=1.0\linewidth]{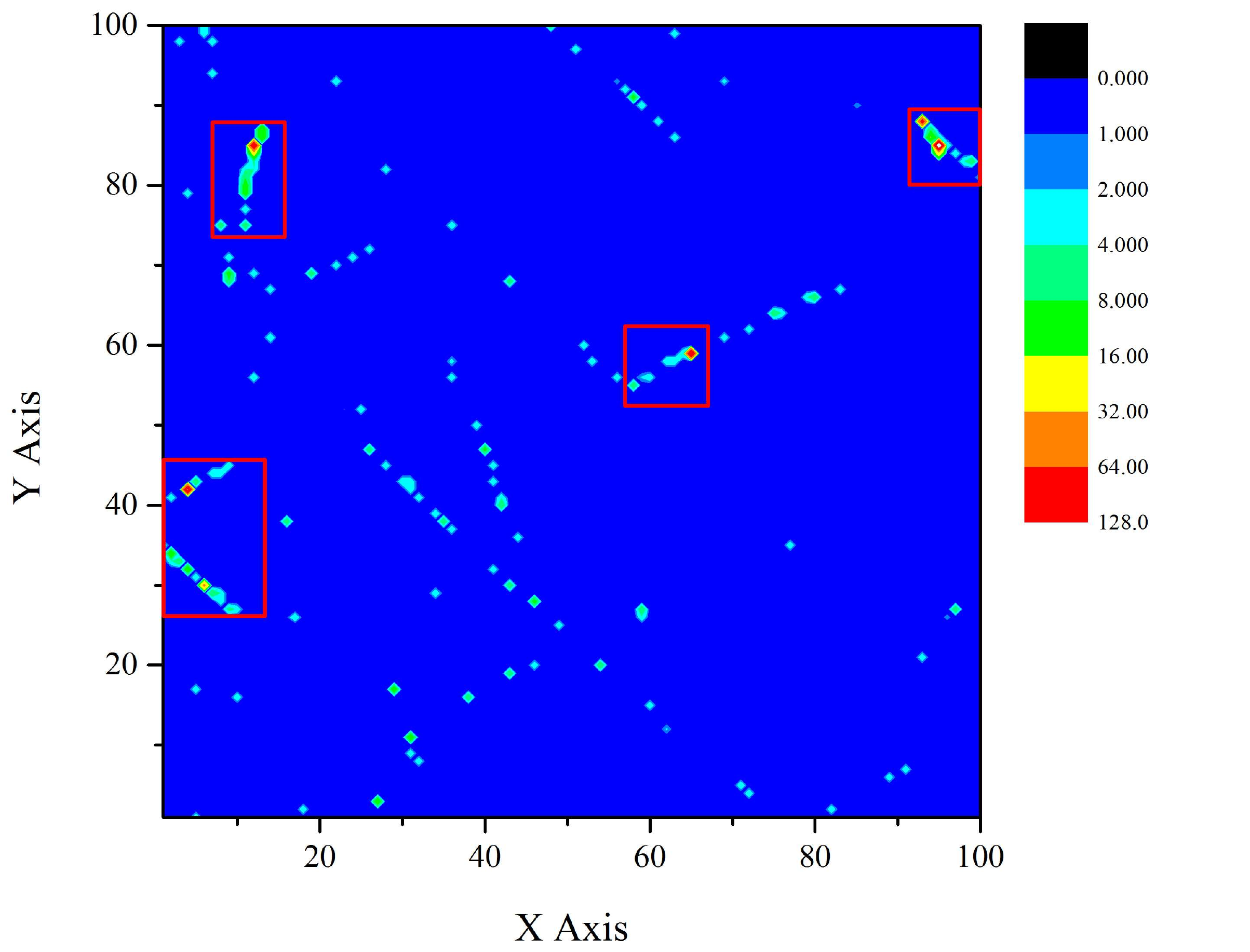}
		\caption{The MU density in the network when $t=36$.}
		\label{fig:MU_dens_32}
	\end{subfigure}
	\caption{The MU density in the monitoring area.}
	\label{fig:MU_dens}
\endminipage\hfill
\minipage{0.32\textwidth}
	\centering
	\begin{subfigure}{1.0\textwidth}
		\centering
		\includegraphics[width=1.0\linewidth]{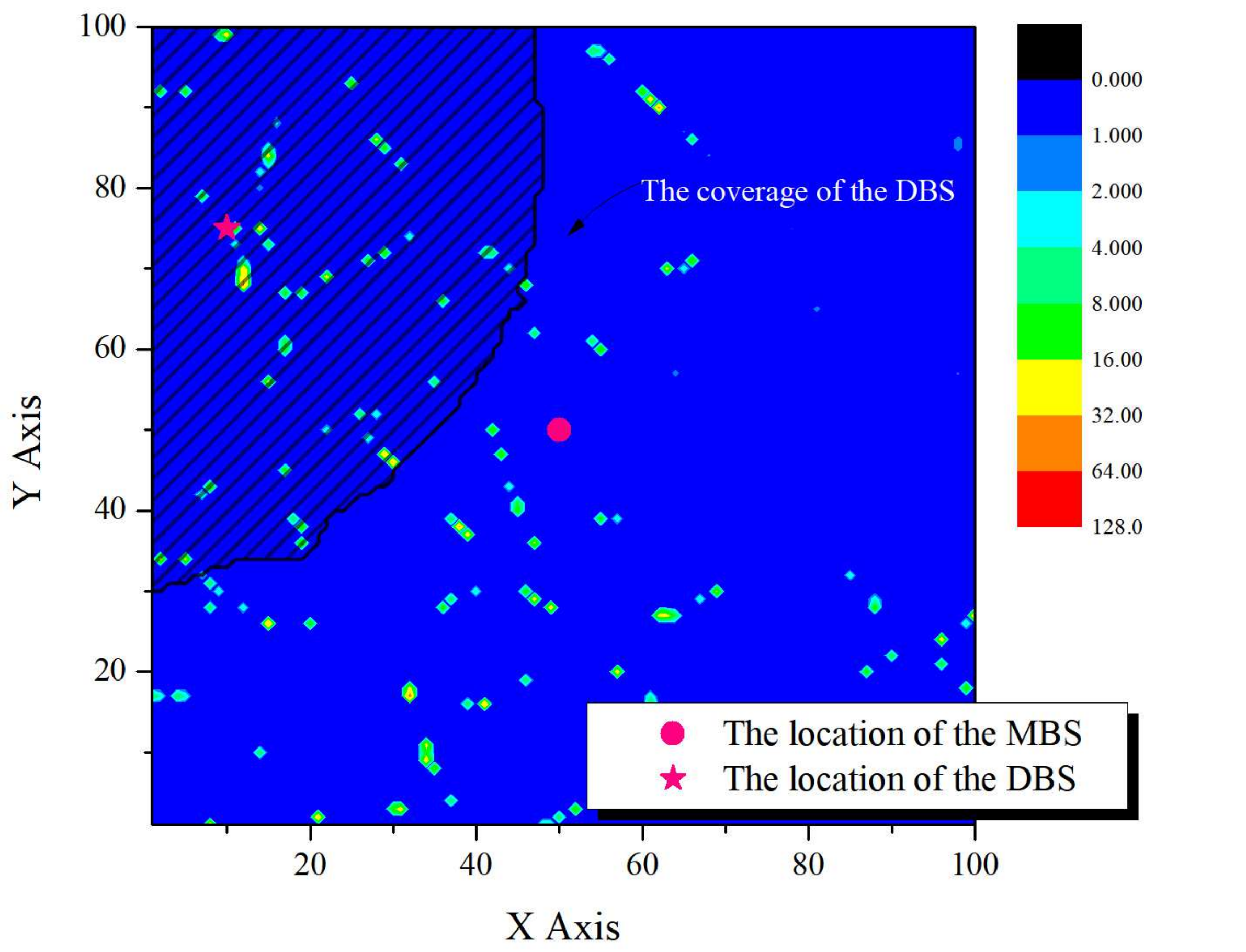}
		\caption{The location and the coverage of the DBS when $t=1$.}
		\label{fig:DBS_placement_1}
	\end{subfigure}%
    \\
	\begin{subfigure}{1.0\textwidth}
		\centering
		\includegraphics[width=1.0\linewidth]{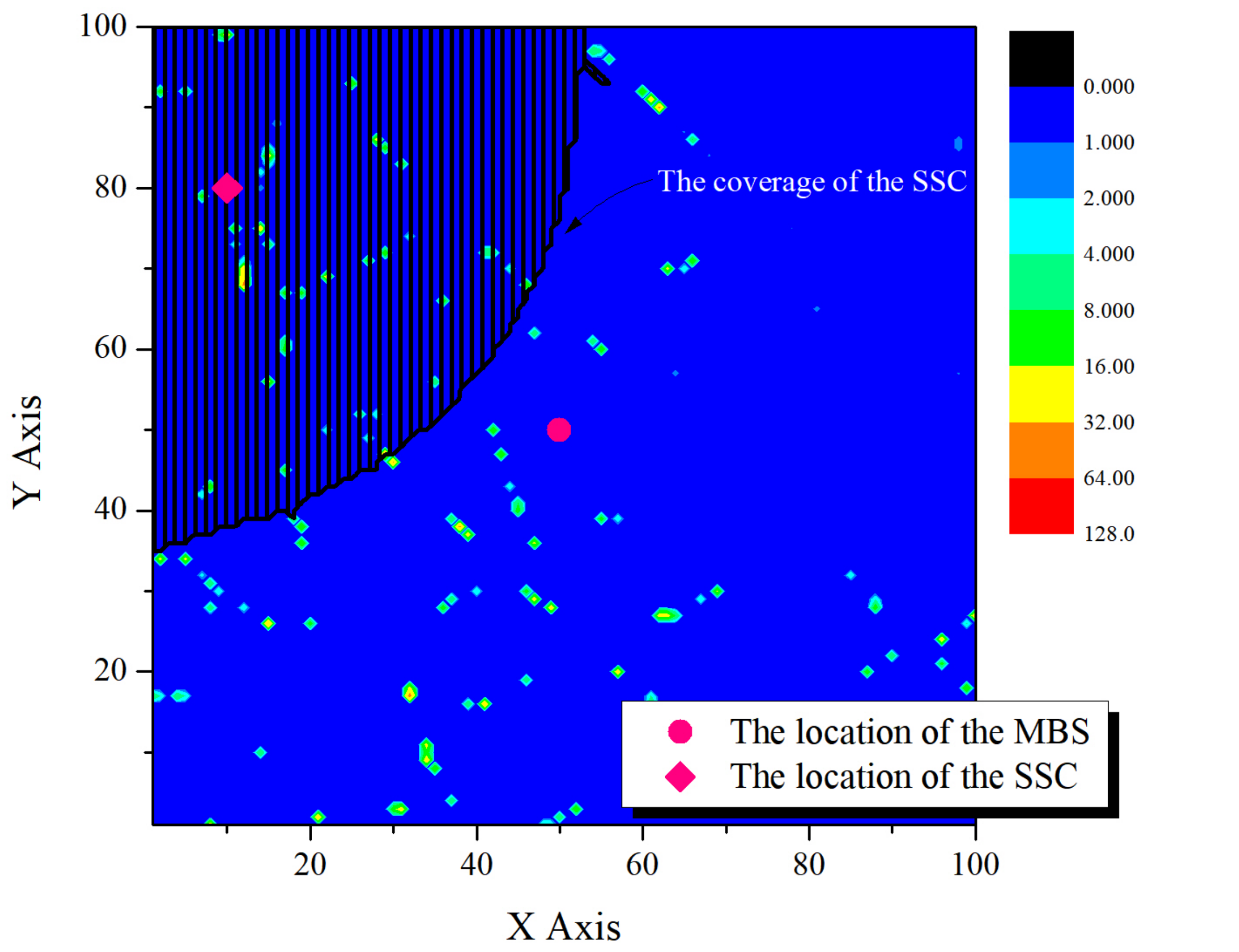}
		\caption{The location and the coverage of the small cell when $t=1$.}
		\label{fig:SSC_placement_1}
	\end{subfigure}
	\caption{The placement of the DBS/small cell when $t=1$.}
	\label{fig:placement_1}
\endminipage\hfill
\minipage{0.32\textwidth}
	\centering
	\begin{subfigure}{1.0\textwidth}
		\centering
		\includegraphics[width=1.0\linewidth]{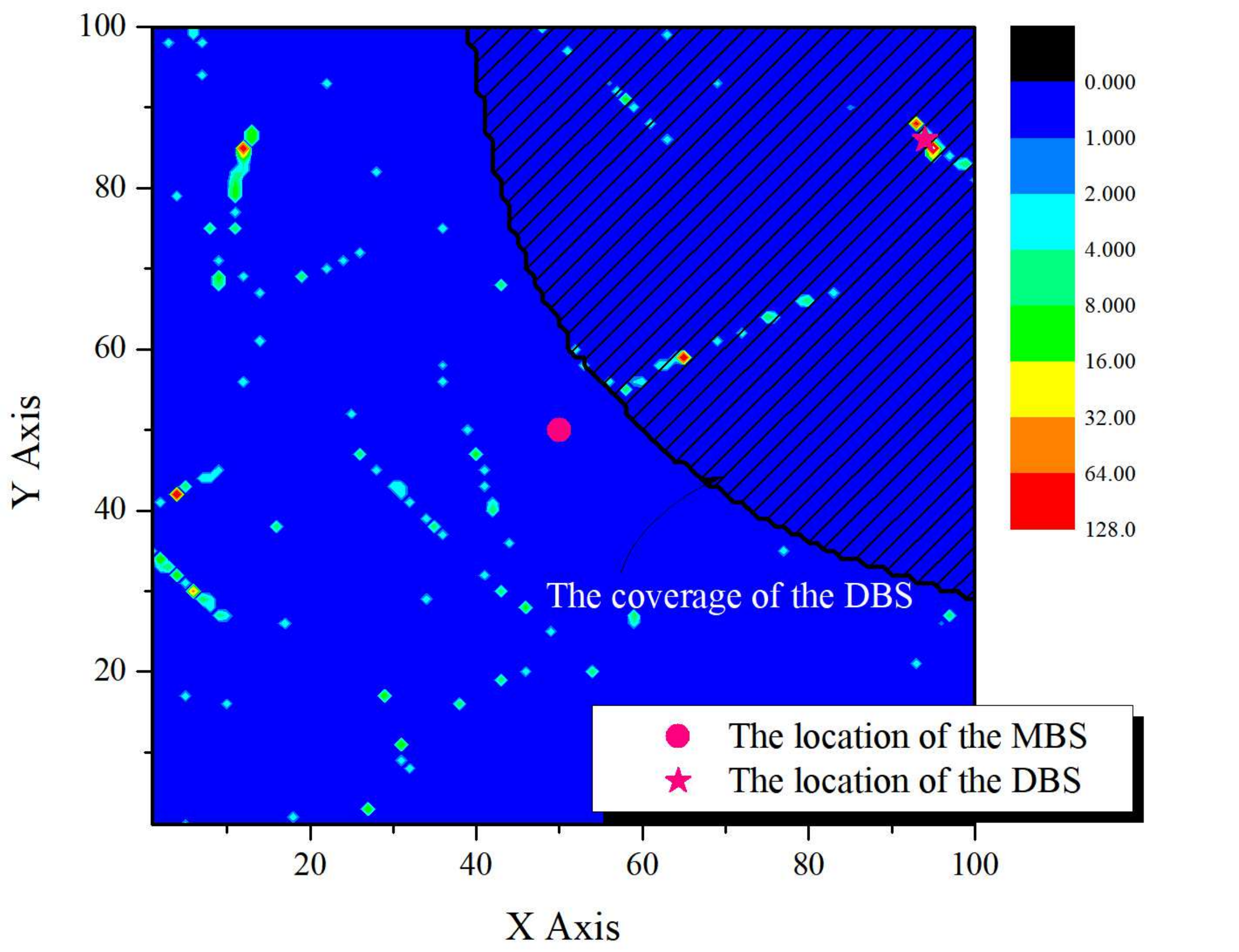}
		\caption{The location and the coverage of the DBS when $t=36$.}
		\label{DBS_placement_36}
	\end{subfigure}%
	\\
	\begin{subfigure}{1.0\textwidth}
		\centering
		\includegraphics[width=1.0\linewidth]{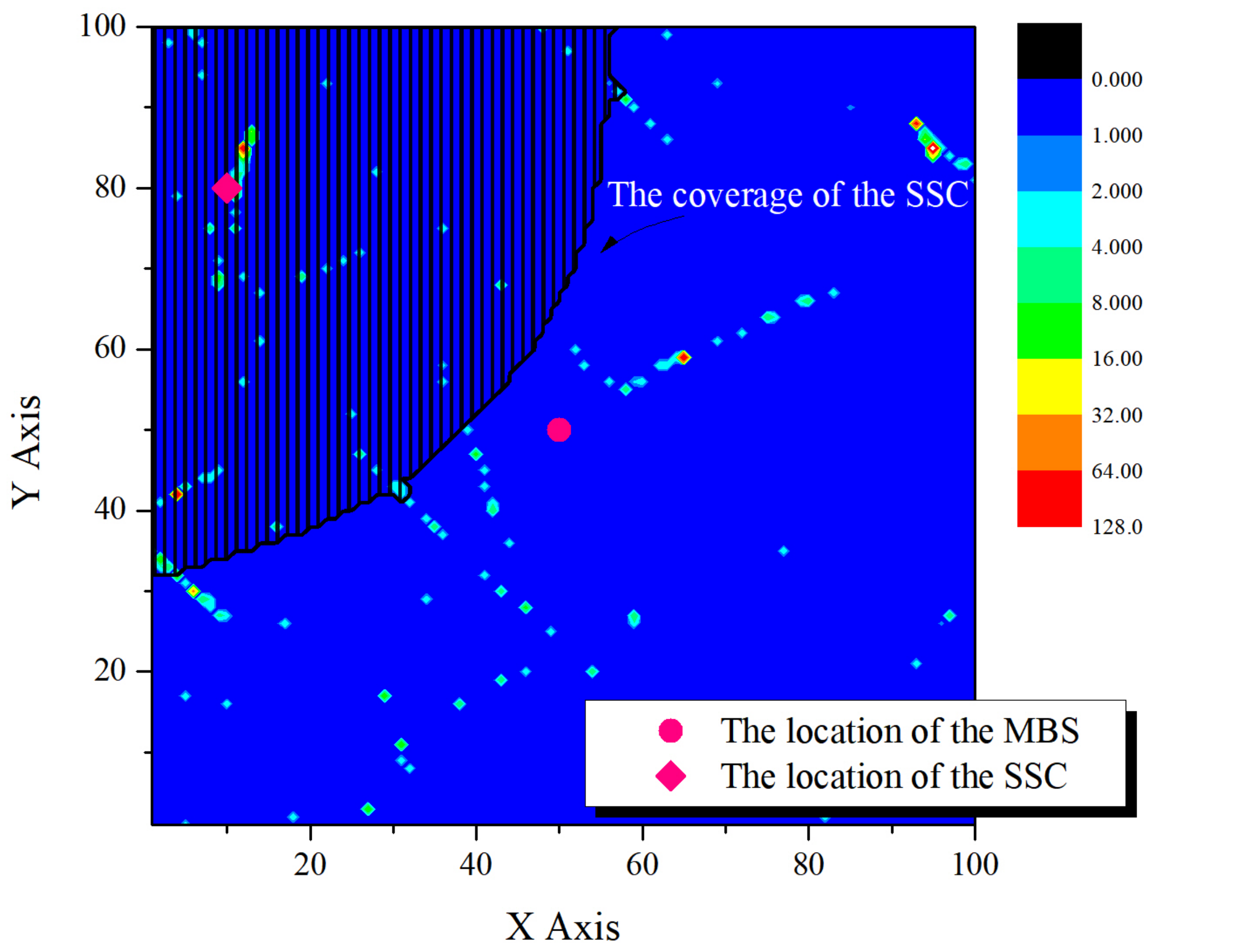}
		\caption{The location and the coverage of the small cell when $t=36$.}
		\label{fig:SSC_placement_36}
	\end{subfigure}
	\caption{The placement of the DBS/small cell when $t=36$.}
	\label{fig:placement_36}
\endminipage\hfill
\end{figure*}

We evaluate the performance of LEAP by comparing it with other two baseline methods, i.e., Single MBS (S-MBS) deployment and Static Small Cell (SSC) deployment. In S-MBS, only one MBS is placed in the area and the MBS uses the whole bandwidth (i.e., 20 $MHz$) to deliver the traffic to MUs. In SSC, one small cell is statically placed in a location to help the MBS in delivering the traffic. We capture the MU density in the monitoring area for the first time slot ($t=1$, i.e., 7 pm--7:10 pm) and the last time slot ($t=36$, i.e., 0:50 am--1:00 am). As shown in Fig. \ref{fig:MU_dens}, the hotspots (which are marked by the red square box) vary in different time slots, but there is one area (i.e., the area around location $<10, 80>$) is the hotspot in both $t=1$ and $t=36$. Thus, we place a static small cell in $<10, 80>$ if SSC is applied.

Fig. \ref{fig:DBS_placement_1} and Fig. \ref{fig:SSC_placement_1} show the placement of the DBS (by applying LEAP) and the static small cell (by applying SSC) when $t=1$, respectively. Clearly, the location and the MU association area of the DBS are very similar with these of the static small cell when $t=1$. However, as shown in Fig. \ref{fig:placement_36}, the DBS is moved to northeast when $t=36$, and thus the location and the MU association area of the DBS are different from those of the static small cell.

\begin{figure}[!htb]
	\centering	
	\includegraphics[width=0.7\columnwidth]{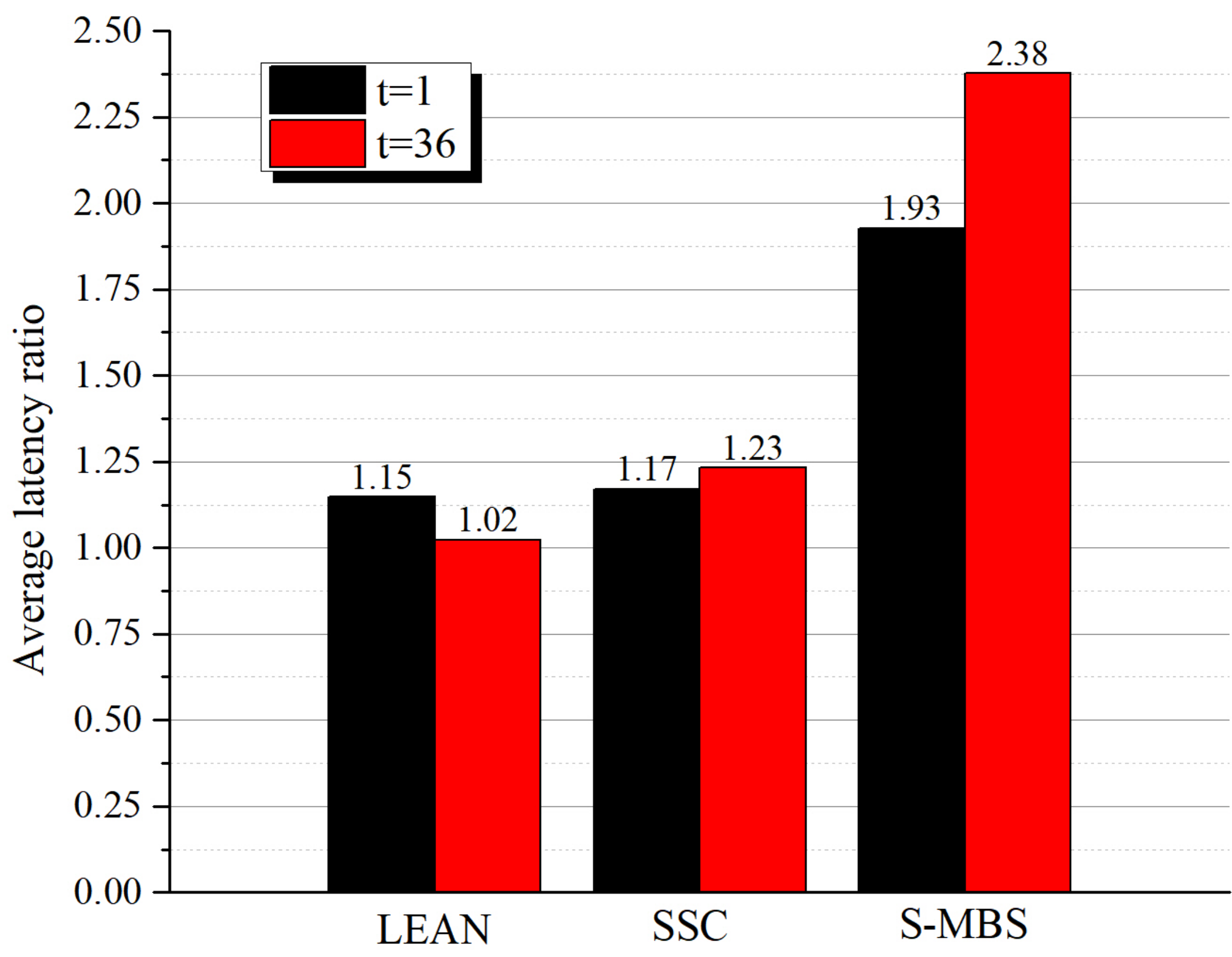}
	\caption{The average latency ratio among MUs.}	
	\label{fig:ave_latency_ratio}
\end{figure}

We further calculate the average latency ratio among the MUs in two different time slots. As shown in Fig. \ref{fig:ave_latency_ratio}, the average latency ratio incurred by LEAP and SSC are similar but much lower than S-MBS when $t=1$. This is because the DBS/static small cell can offload the traffic loads from the MBS such that the average waiting time of the MUs is reduced. When $t=36$, the average latency ratio incurred by LEAP is lower than SSC. This is because a new hotspot appears in the network when $t=36$ and the previous location is not the best choice to minimize the average latency ratio of MUs. Consequently, the DBS is moved to a better location and optimizes its association coverage to reduce the average latency ratio. Hence, we conclude that LEAP can automatically optimize the location and the association coverage of the DBS to reduce the average latency ratio of MUs in each time slot. 

\section{Conclusion}
In this paper, we investigate the DBS placement problem in a heterogeneous network by considering the energy consumption limitation of the DBS. We consider the QoS of an MU as the latency ratio of the MU. We formulate the DBS placement problem as an optimization problem to minimize the total average latency ratio among the MUs. We design the LEAP algorithm to solve the optimization problem and demonstrate the performance of LEAP via simulations.

\ifCLASSOPTIONcaptionsoff
  \newpage
\fi


\begin{thebibliography}{1}

\bibitem{Sedjelmaci:IDE:2017}
H. Sedjelmaci, S. M. Senouci, and N. Ansari, ``Intrusion Detection and Ejection Framework Against Lethal Attacks in UAV-Aided Networks: A Bayesian Game-Theoretic Methodology,'' in \emph{IEEE Transactions on Intelligent Transportation Systems}, vol. 18, no. 5, pp. 1143-1153, 2017.

\bibitem{Sun:MEC:2017}
X. Sun and N. Ansari, ``Mobile Edge Computing Empowers Internet of Things,'' NJIT Advanced Networking Lab., Tech. Rep. TR-ANL-2017-003; also archived in Computing Research Repository (CoRR), arXiv:1709.00462, 2017.

\bibitem{Nokia}
Imran Khan Durrani, ``Nokia and EE – pushing the limits of technology with drones.'' [Online]. Available: https://blog.networks.nokia.com/partners-and-customers/2016/08/11/nokia-ee-pushing-limits-technology-drones/

\bibitem{Sharma:2016:UAH}
V. Sharma, M. Bennis, and R. Kumar, ``UAV-Assisted Heterogeneous Networks for Capacity Enhancement,'' in \emph{IEEE Communications Letters}, vol. 20, no. 6, pp. 1207--1210, June 2016.

\bibitem{Sun:2015:EBA}
X. Sun and N. Ansari, ``Energy-optimized bandwidth allocation strategy for mobile cloud computing in LTE networks,'' in \emph{2015 IEEE Wireless Communications and Networking Conference (WCNC)}, New Orleans, LA, 2015, pp. 2120--2125.

\bibitem{Zeng:2016:WCU}
Y. Zeng, R. Zhang, and T. J. Lim, ``Wireless communications with unmanned aerial vehicles: opportunities and challenges,'' in \emph{IEEE Communications Magazine}, vol. 54, no. 5, pp. 36--42, May 2016.

\bibitem{AlHourani:2014:OLA}
A. Al-Hourani, S. Kandeepan, and S. Lardner, ``Optimal LAP Altitude for Maximum Coverage,'' in \emph{IEEE Wireless Communications Letters}, vol. 3, no. 6, pp. 569--572, Dec. 2014.

\bibitem{Mozaffari:2015:DSC}
M. Mozaffari, W. Saad, M. Bennis, and M. Debbah, ``Drone Small Cells in the Clouds: Design, Deployment and Performance Analysis,'' in \emph{2015 IEEE Global Communications Conference (GLOBECOM)}, San Diego, CA, 2015, pp. 1--6.

\bibitem{Yaliniz:2016:EPA}
R. Yaliniz, A. El-Keyi, and H. Yanikomeroglu, ``Efficient 3-D placement of an aerial base station in next generation cellular networks,'' \emph{arXiv preprint arXiv:1603.00300}, 2016.

\bibitem{Fotouhi:2016:DBS}
A. Fotouhi, M. Ding, and M. Hassan, ``Dynamic Base Station Repositioning to Improve Performance of Drone Small Cells,'' in \emph{2016 IEEE Globecom Workshops (GC Wkshps)}, Washington, DC, 2016, pp. 1--6.

\bibitem{Kleinrock:1976:QSC}
L. Kleinrock, \emph{Queueing Systems: Computer Applications}, vol. 2. New York, NY USA: Wiley, 1976.

\bibitem{Han:2016:TLB}
T. Han and N. Ansari, "A Traffic Load Balancing Framework for Software-Defined Radio Access Networks Powered by Hybrid Energy Sources," in \emph{IEEE/ACM Transactions on Networking}, vol. 24, no. 2, pp. 1038-1051, April 2016.

\bibitem{3GPP:2012:release_11}
3GPP, ``3GPP TR 36.828 version 11.0.0, release 11: 3rd generation partnership project; further enhancements to LTE time division duplex (TDD) for downlink-uplink (DL-UL) interference management and traffic adaptation,'' \emph{3GPP Technical Report}, Tech. Rep., 2012.

\bibitem{user_trace}
User Movement Simulations Project. Available. [Online]: http://everywarelab.di.unimi.it/lbs-datasim

\bibitem{Franco:2015:ECP}
C. D. Franco and G. Buttazzo, ``Energy-Aware Coverage Path Planning of UAVs,'' in \emph{2015 IEEE International Conference on Autonomous Robot Systems and Competitions}, Vila Real, 2015, pp. 111--117.

\bibitem{Auer:2011:HME}
G. Auer, \emph{et al.}, ``How much energy is needed to run a wireless network?,'' in \emph{IEEE Wireless Communications}, vol. 18, no. 5, pp. 40--49, Oct. 2011.

\end{thebibliography}
\end{document}